\documentclass[showpacs, pra,twocolumn,preprintnumbers ,amsmath, amssymb, superscriptaddress, aps]{revtex4-2}

\usepackage[utf8x]{inputenc}
\usepackage{color}
\usepackage{amsmath,amssymb}
\usepackage{pifont}% http://ctan.org/pkg/pifont
\usepackage{amssymb}  % More symbols
\usepackage{bbold}
\usepackage{float}
\usepackage{subfloat}

\usepackage[caption=false]{subfig}
\usepackage{tikz}
\usepackage{makecell}
\usepackage{subfig}
\usepackage{pifont}   % Ding symbols
\usepackage{graphicx} % Include figure files
\graphicspath{{Figures/}}
\usepackage{dcolumn}  % Align table columns on decimal point
\usepackage{bm}       % bold math
\usepackage{multirow} % Table functions
\usepackage{placeins}% For making floats not move around everywhere
\usepackage[colorlinks]{hyperref}
\usepackage{mathtools}
\usepackage{appendix}

\def \be{\begin{align}}
	\def \ee{\end{align}}
\def \bea{\begin{eqnarray}}
	\def \eea{\end{eqnarray}}

\begin{document}

	\title{Laser-assisted tunneling in a static tungsten  diselenide
		WSe$_2$ barrier}
	
	\author{Rachid El Aitouni}
	
	\affiliation{Laboratory of Theoretical Physics, Faculty of Sciences, Choua\"ib Doukkali University, PO Box 20, 24000 El Jadida, Morocco}
	
	\author{Mohammed El Azar}
	\affiliation{Laboratory of Theoretical Physics, Faculty of Sciences, Choua\"ib Doukkali University, PO Box 20, 24000 El Jadida, Morocco}
	
	\author{Clarence Cortes}
	\affiliation{Vicerrector\'ia de Investigaci\'on y Postgrado, Universidad de La Serena, La Serena 1700000, Chile}
	\author{Pablo Díaz}
	\affiliation{Departamento de Ciencias F\'{i}sicas, Universidad de La Frontera, Casilla 54-D, Temuco 4811230, Chile}  
	\author{David Laroze}
	\affiliation{Instituto de Alta Investigación, Universidad de Tarapacá, Casilla 7D, Arica, Chile}
	\author{Ahmed Jellal}
	\email{a.jellal@ucd.ac.ma}
	\affiliation{Laboratory of Theoretical Physics, Faculty of Sciences, Choua\"ib Doukkali University, PO Box 20, 24000 El Jadida, Morocco}
	
	\begin{abstract}

			We study the tunneling effect of Dirac fermions in a monolayer WSe$_2$ subjected to a static electrostatic barrier and irradiated by a linearly polarized laser field. Within the Floquet formalism, the time-periodic driving is incorporated to derive analytical wave functions across the three regions of the system. By enforcing continuity conditions at the interfaces, we obtain the transmission and reflection coefficients, which are then used to evaluate the conductance via the Büttiker approach. Our results reveal that the laser field induces a rich Floquet sideband structure, whose number and strength increase with the driving parameter $\alpha$. This leads to a significant suppression of transmission and provides an efficient mechanism to overcome Klein tunneling. Moreover, increasing the width of the irradiated region enhances the interaction between fermions and the external field, resulting in energy renormalization and the formation of Stark-like confined states. The interaction between several Floquet channels creates strong interference effects, which reduce the transmitted current even further. The results demonstrate that light-matter interaction allows for the dynamic control of quantum transport in WSe$_2$ materials. This technology allows for the development of new optoelectronic devices, including tunable quantum filters and light-controlled nanoscale transistors.
		
	\end{abstract}
	
	\maketitle
	
	\section{Introduction}	\label{Intro}
	For a long time, technological development has been oriented toward the search for new materials with extremely small dimensions capable of delivering superior performance. The isolation of a single graphene sheet in 2004 by the Manchester research group marked a major breakthrough \cite{Novos2004}. This two-dimensional (2D) material are defined in general as having a thickness below one nanometer \cite{Novoselov2016}. They quickly became a revolutionary candidate in advanced technologies due to its exceptional properties, such as flexibility, rigidity, high electronic mobility, electron velocity (approximately 300 times smaller than the speed of light \cite{mobilfermgrap1,mobilfermgrap2}) and remarkable transparency, absorbing only $2.3\%$ incident light \cite{propgraph,absorgrap}. Nevertheless, despite these extraordinary features, the use of graphene in electronics remains limited because of the difficulty in confining its fermions \cite{zerogap1,zerogap2,gapgraph1,gapgraph2}. Several approaches have been proposed to open an energy gap---mechanical deformation \cite{deformation1,deformation2}, deposition on substrates \cite{Gapsubst,substrateffet}, stacking of graphene layers, or graphene doping \cite{dopagedegraphene}, but these methods remain insufficient. Even when a gap is introduced, another challenge arises: Klein tunneling \cite{klienexp,klien1,klien2}. Researchers have suggested possible solutions, such as applying static or time-dependent (oscillating) barriers \cite{potentieloscil1,potentieloscil2,potentieloscil22,timepot2,doublebarriertemps,magetic1,magetic4,magnetic3,magneticfield,biswasgraphmagnet,Elaitouni2023,doublemagetic,doublelasermagn2,doublelaser,triangbarmag}. 
		% Light irradiation of graphene generally creates Floquet modes, since the Hamiltonian in this case becomes time‑periodic. As a result, the energy spectrum is quantized, leading to the formation of a Floquet‑induced gap, which has been experimentally observed in 2024 \cite{Das24}, as well as the emergence of dynamical bound states \cite{Gregefalk23}. Polarization strongly affects tunneling \cite{2012rev} and also conductance \cite{Iurov2019}. Moreover, such irradiation can partially suppress Klein tunneling 
		Light irradiation of graphene generally gives rise to Floquet modes, as the Hamiltonian becomes time-periodic. Consequently, the energy spectrum becomes quantized, leading to the formation of a Floquet-induced gap, which has been experimentally observed \cite{Das24}, as well as the emergence of dynamical bound states \cite{Gregefalk23}. The polarization of the applied field plays a crucial role, strongly influencing both tunneling processes \cite{2012rev} and electronic conductance \cite{Iurov2019}. Moreover, such irradiation can partially suppress Klein tunneling and lead to polarization-dependent transport behavior
		\cite{Iurov2012,Iurov2020}.

	% Following the success of graphene, new materials have been explored for technological development, particularly transition metal dichalcogenides (TMDs) \cite{valleytrodichlo,optopropDichloc,Chloappl}. These 2D materials consist of two chalcogen atoms (S, Se, Te $\cdots$) bonded to transition metals from groups IV, V, or VI. Their layers are held together by weak van der Waals forces \cite{vander}, which makes them easy to isolate. Molybdenum disulfide (MoS$_2$) was first exfoliated in 2005, tungsten diselenide (WSe$_2$) in 2013, and molybdenum diselenide (MoSe$_2$) in 2015  \cite{vander}. Unlike graphene, these materials exhibit strong spin--orbit coupling (SOC) \cite{OSCgraph}, opening promising perspectives in advanced technologies, particularly in spintronics and valleytronics \cite{Spinovalley,gs,gv}. Although these materials are semiconductors with a nonzero bandgap, current research aims to further develop and optimize their physical properties. 
	
	Following the success of graphene, new materials have been explored for technological development, particularly transition metal dichalcogenides (TMDs) \cite{valleytrodichlo,optopropDichloc,Chloappl}.
	The 2D materials display two chalcogen atoms (S, Se, Te $\cdots$), which bond with transition metals from groups IV, V, or VI. The material layers become easy to separate because the van der Waals forces which connect them throughout the material exist between the individual layers of the material \cite{vander}. The first time researchers separated molybdenum disulfide (MoS$_2$) from other materials occurred in 2005, while they achieved tungsten diselenide (WSe$_2$) separation in 2013 and molybdenum diselenide (MoSe$_2$) separation in 2015 \cite{vander}. It is discovered that these materials create strong spin–orbit coupling (SOC) which enables development of advanced technologies for both spintronics and valleytronics applications \cite{OSCgraph,2Dappl,grapappl,2Dappl2,Spinovalley,gs,gv}. The materials function as semiconductors because they maintain a bandgap, yet researchers continue to investigate their physical characteristics to achieve better results.

	We theoretically investigate the behavior of fermions in tungsten diselenide under the influence of a static potential barrier of height $V_0$ applied over a region of width $D$, irradiated by a linearly polarized laser field of amplitude $A_0$ and frequency $\omega$. The time-dependent oscillation of the Hamiltonian requires the use of Floquet theory \cite{Floquet} to determine the spinors in the three regions of the proposed structure. The continuity of the wave function across these regions leads to four equations, each involving an infinite number of Floquet modes, which complicates the resolution and necessitates a matrix formalism. The continuity equation is used to express the three current densities, from which transmission and reflection probabilities are derived. To connect microscopic quantities to macroscopic observables, the Büttiker formalism is employed to express the conductance \cite{conduct2,butker,conduct1}. Numerical results demonstrate that laser irradiation modifies tunneling by redistributing the transmission probability among the Floquet sidebands and by changing whether the corresponding modes inside the barrier are propagating or evanescent.
	% laser irradiation becomes a crucial tool for reducing transmission while enabling the channeling of fermion transport across the barrier.
	The photon exchange between the barrier and the fermions modifies their energy, which on the one hand traps the fermions in new energy levels (known as the Stark effect \cite{Stark}), and on the other hand leads to the emergence of new transmission channels. Increasing the laser field intensity suppresses Klein tunneling for most incidence angles agreement with the same experiment in graphene \cite{Iurov2012,Iurov2020}. Transmission without photon exchange proves to be more probable than other Floquet modes. Conductance follows the behavior of transmission and decreases with increasing laser intensity. Moreover, enlarging the barrier width enhances the interaction of fermions with the laser field and strengthens destructive interference, thereby reducing the probability of transmission.

	Several studies have investigated tunneling through oscillating barriers in Weyl 
		semimetals, focusing on how fermions interact with time-dependent potentials 
		\cite{Das24,Gregefalk23}. Other works have examined the interaction of light 
		with fermions tunneling through gapless systems such as graphene 
		\cite{Iurov2022,Kristinsson2016,Iurov2012}, Dirac \cite{IurovPRR2020}, and 
		Weyl semimetals, showing that light with linear or circular polarization can 
		open topological gaps, generate Floquet resonances, and modify both transmission 
		and conductance. In contrast, the present work explores a distinct regime: a 
		semiconductor WSe$_2$ \cite{Li} with a large band gap subjected to linearly 
		polarized laser irradiation, where the dominant effect is not valley asymmetry 
		but rather the symmetric suppression of Klein tunneling through dynamic 
		localization. The intrinsic gap of WSe$_2$ acts as a natural confining barrier, 
		allowing fermions to be trapped without requiring additional external constraints, 
		unlike gapless systems such as graphene \cite{Kristinsson2016} and Weyl 
		semimetals \cite{Das24,Gregefalk23}, which require extra symmetry breaking for 
		carrier control. Moreover, the strong spin--orbit coupling in WSe$_2$ makes its 
		conductance particularly sensitive to external fields \cite{Li,Yang2024}, since 
		it induces a significant band splitting that directly couples to the 
		laser-induced modulation.
	
	%To the best of our knowledge, the demonstration of dynamic localization and the consequent suppression of Klein tunneling driven by a vector potential in a strongly spin-orbit coupled, massive Dirac material like WSe$_2$ has not been reported in the existing literature.}

The paper is organized as follows. In Sec. \ref{Theory}, we present the theoretical model and derive the wave functions in the different regions of the system. This section establishes the fundamental framework needed to describe the tunneling process through the WSe$_2$ barrier. Based on the analytical results derived in Appendices \ref{Tran} and \ref{Cond}, Sec. \ref{num} is devoted to a comprehensive numerical analysis of the transmission probability and conductance. We systematically investigate how variations in the system parameters affect the transport properties of the WSe$_2$ barrier. Special attention is given to the role of laser-assisted processes and their impact on the electronic response of the system. Finally, Sec. \ref{concl} provides a summary of the main results and presents the overall conclusions of the study.

%    The paper is organized as follows. In Sec. \ref{Theory}, we present the theoretical model and determine the wave functions in the different regions of the system. Based on the analytical results derived in Appendices \ref{Tran} and \ref{Cond}, Sec. \ref{num} is devoted to the numerical analysis of the transmission and conductance, where we examine the effects of varying the system parameters on the properties of the WSe$_2$ barrier. Finally, Sec. \ref{concl} provides summary and  conclusions.

%The paper is organized as follows. In Sec. \ref{Theory}, we present the theoretical model and determine the wave functions corresponding to each region. In Sec. \ref{transm}, the boundary conditions are applied to obtain the transmission, which then leads to determining the conductance by using the Büttiker relation. In Sec. \ref{num}, we present our numerical results to analyze the effect of varying the parameters of our system on the properties of WSe$_2$. Finally, Sec. \ref{concl} provides a summary and  conclusion. 

\section{Theoritical Model}	\label{Theory}
\begin{figure}[H]
	\centering
	\includegraphics[scale=0.23]{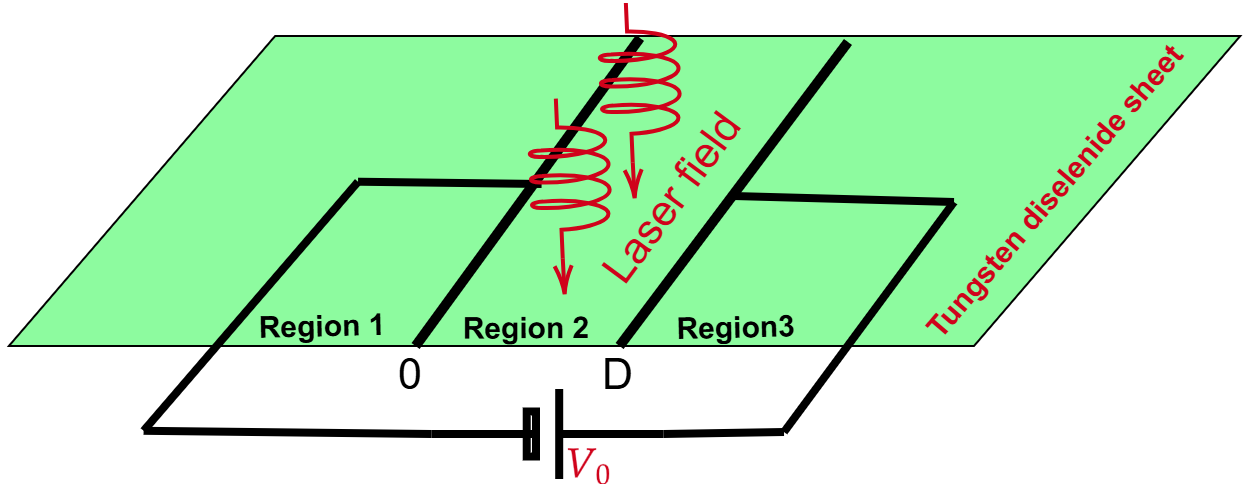}
	\caption{The schematic illustrates a static barrier of height $V_0$ applied over a region of width $D$ on a WSe$_2$ sheet, where the region is irradiated by a linearly polarized laser field.
	}\label{str}
\end{figure}

Tungsten diselenide (WSe$_2$) is theoretically investigated. %The proposed configuration involves the irradiation of a region of width $D$ by a linearly polarized laser field with amplitude $A_0$ and frequency $\omega$.  This laser field originates from a sinusoidal electric field. As a result, the irradiation divides the WSe$_2$ sheet into three distinct regions, as illustrated in Fig.  \ref{str}.
The proposed configuration can be realized experimentally as follows.
	Two rectangular  electrodes are placed perpendicularly to the WSe$_2$ sheet. 
	Connected to a voltage generator $V_0$ \cite{klienexp}, these electrodes create a static potential barrier of height $V_0$. 
	The region between the two electrodes is then subjected to linearly polarized laser irradiation 
	with amplitude $A_0$ and frequency $\omega$. 
	The laser field is generated by a sinusoidal electric field \cite{Sinha2012}.  This configuration
	divides the sheet into three distinct regions, 
	as illustrated in Fig. \ref{str}.
%{\bb We model the linearly polarized laser field using a macroscopic time-dependent vector potential $\vec{A}(t) = A_0 \cos(\omega t)\,\hat{e}_x$ directed along the transport axis. The light-matter interaction is introduced into the bare Hamiltonian $H_0$ via the minimal coupling principle (Peierls substitution), replacing the momentum operator $\vec{p}$ with $\vec{p} + e\vec{A}(t)$. Consequently, the total time-dependent Hamiltonian becomes:
	%\begin{equation}    H(t) = H_0(\vec{p}) + \tau\, e v_F A_0 \cos(\omega t)\, \sigma_x,\end{equation}
	%where the second term represents the periodic radiation-induced interaction $H_{\mathrm{int}}(t)$ that couples the energy states and generates the Floquet sidebands.}

	The effect of linearly polarized radiation is incorporated using the standard minimal coupling (Peierls substitution), where the canonical momentum is replaced as $\mathbf{p} \rightarrow \mathbf{p} + e\mathbf{A}(t)$. We consider an electromagnetic wave propagating perpendicular to the sample and linearly polarized along the $y$-direction. Within the dipole approximation \cite{dipolapprox}, the electric field is given by $E(t)=F\sin(\omega t)$, leading to the vector potential
	\begin{equation}
		\mathbf{A}(t) = (0, A_0 \cos(\omega t), 0)
	\end{equation}
	where $A_0 = F/\omega$ is the amplitude and $\omega$ is the frequency. This choice implies that only the $y$-component of the momentum is modified, i.e., $p_y \rightarrow p_y + eA_0 \cos(\omega t)$.
	As a consequence, within the low-energy model \cite{low,WSe2valeurs}, the Hamiltonian becomes explicitly time-dependent and can be written as
	\begin{equation}
		H = H_0 + H_L
	\end{equation}
	where $H_0$ is the intrinsic (static) Hamiltonian and $H_L$ describes the interaction with the radiation field. These terms are given by
	\begin{widetext}
		\begin{align}
			H_0 &= v_F \left[ \tau \sigma_x p_x + \sigma_y p_y \right] + \frac{\Delta}{2}\sigma_z 
			-\lambda_c \tau s_z\frac{(\sigma_0+\sigma_z)}{2} 
			-\lambda_v \tau s_z\frac{(\sigma_0-\sigma_z)}{2}
			+ V_0\sigma_0 \\
			H_L &= v_F \left[\tau\sigma_x A_{t_x} + \sigma_y A_{t_y} \right].
		\end{align}
	\end{widetext}
	Here, $p_{x,y}$ are the momentum components, $\sigma_i$ ($i=x,y,z$) are the Pauli matrices, $\tau=\pm1$ labels the ${K}$ and ${K'}$ valleys, and $s_z=\pm1$ denotes spin up and down. The material parameters are $v_F=5\times10^5$ m/s, $\Delta=1.7$ eV, and the spin--orbit couplings are $\lambda_v=112.5$ meV and $\lambda_c=7.5$ meV \cite{WSe2valeurs}.
	%	Using the above form of the vector potential, the Hamiltonian can be equivalently written as
	%	\begin{widetext}
		%		\begin{align}
			%			H= v_F \left[ \tau \sigma_x p_x + \sigma_y \left(p_y + eA_0 \cos(\omega t)\right) \right] 
			%			+ \frac{\Delta}{2}\sigma_z 
			%			-\lambda_c \tau s_z\frac{(\sigma_0+\sigma_z)}{2} 
			%			-\lambda_v \tau s_z\frac{(\sigma_0-\sigma_z)}{2}
			%			+ V_0\sigma_0.
			%		\end{align}
		%	\end{widetext}
	The time-periodic Hamiltonian $H$ captures the light--matter interaction and leads to photon-assisted transport processes, which are treated using the Floquet formalism in the following.
		We can write this Hamiltonian in the matrix form as 
		\begin{align}\label{Ham2}
			H =
			\begin{pmatrix}
				V_0+	\frac{\Delta}{2} - \lambda_c \tau s_z & v_F \left[\tau p_x - i(p_y+eA_t)\right] \\
				v_F \left[\tau p_x + i(p_y+eA_t)\right] & V_0-\frac{\Delta}{2} +\lambda_v \tau s_z
			\end{pmatrix}.
		\end{align}
		
		The time-dependent oscillation of the field generates Floquet states. However, the wave function is time-dependent,  and at each instant states are either created or annihilated. 
		The motion of fermions is studied only along the $x$ direction, and then the momentum $k_y$ is conserved.  Consequently, the spinors associated with \eqref{Ham2} take the form
		$\Psi^j(x,y,t) = \binom{\psi_{c}^j(x)}{\psi_{v}^j(x)}\,\phi(t)e^{ik_yy}\,e^{-iE t/\hbar}$.
		To obtain the spinors for each valley $\tau$ and spin $s_z$, one  solve the  eigenvalue equation $H \Psi^j(x,y,t)= E \Psi^j(x,y,t)$. This process yields  two coupled equations
		\begin{widetext}
			\begin{align}
				&	\left[i\hbar \dfrac{\partial}{\partial t} -
				\left(V_0+\frac{\Delta}{2} -\lambda_c \tau s_z\right)\right]
				\psi_c(x)\phi(t)e^{-iEt/\hbar}{e^{ik_yy}}=
				v_F \left[-i\hbar \tau \partial_x {- \hbar \partial_y} - i eA_0 \cos(\omega t)\right]\psi_v(x)\phi(t)e^{-iEt/\hbar} {e^{ik_yy}}\\
				&	\left[	i\hbar \frac{\partial}{\partial t}
				-\left(V_0-\frac{\Delta}{2} + \lambda_v \tau s_z\right)\right]
				\psi_v(x)\phi(t)e^{-iEt/\hbar} {e^{ik_yy}}=
				v_F \left[-i\hbar \tau \partial_x{+\hbar \partial_y}+ i eA_0 \cos(\omega t)\right]\psi_c(x)\phi(t)e^{-iEt/\hbar}{e^{ik_yy}}.
			\end{align}
		\end{widetext}
		To determine $\phi(t)$, we assume that the eigenvalue equations hold in the absence of the laser. Hence, following \cite{laser1,laser2,laser3}, the function $\phi(t)$ can be expressed as
		\begin{align}
			\phi(t)=e^{-i\frac{A_0}{\omega}\sin(\omega t)}=\sum_{-\infty}^{+\infty}J_m(\alpha)e^{-im\omega t}
		\end{align}
		where  $A_0\cos(\omega t)=\partial_t e^{-i\frac{A_0}{\omega}\sin(\omega t)}$,  $J_m(\alpha)$ is the function of Bessel,   $m$ the Floquet mode, and we have defined the parameter $\alpha=\frac{A_0}{\omega}$. As a result, eigenspinor in each region can be written as
		\begin{equation}
			\Psi^j(x,y,t)=\sum_{m=-\infty}^{\infty}\binom{\psi_{c}^j(x)}{\psi_{v}^j(x)}J_m(\alpha)e^{ik_yy}e^{-i(E+m\omega \hbar) t/\hbar}
		\end{equation}
		Then the eigenvalue equations can be expressed as 
		%\begin{widetext}
		\begin{align}
			&\left[\Delta_c - (E+m\hbar\omega)\right]\phi_c -i \hbar v_F \left(\tau \partial_x +k_y+\frac{m \omega}{v_F}\right)\phi_v=0\\
			& i \hbar v_F \left(-\tau \partial_x +k_y+\frac{m \omega}{v_F}\right)
			\phi_c+\left[\Delta_v - (E+m\hbar\omega)\right]\phi_v=0
		\end{align}
		%\end{widetext}
		and we have set
		\begin{align}
			&   \Delta_c=V_0+\frac{\Delta}{2} -\lambda_c \tau s_z\\
			&  \Delta_v=V_0-\frac{\Delta}{2} +\lambda_v \tau s_z.
		\end{align}
		{In  region 2 where a laser field is applied, the eigenspinor %ithe eigenspinor 
			for a given valley $\tau$ takes the form}
		%\begin{widetext}
			\begin{align}\label{psi2}
				\Psi^2(x,y,t)=&\sum_{m,l-\infty}^{\infty}\left[a^m\begin{pmatrix}
					1\\ \Gamma_m
				\end{pmatrix}e^{i\tau q^m_xx}+b^m\begin{pmatrix}
					1\\ -\Gamma^*_m
				\end{pmatrix}e^{-i\tau q^m_xx}\right]\notag\\
				&e^{ik_yy}e^{-i(E+m\hbar\omega)t/\hbar}J_{l-m}(\alpha)
			\end{align}
		%\end{widetext}  
		where the vave vector component and complex number are given by
		\begin{widetext}
			\begin{align}
				&	(q^m_x)^2 =\frac{(E+m\hbar\omega-\Delta_c)(E+m\hbar\omega-\Delta_v)}{(\hbar v_F)^2}-(k_y+m\omega/v_F)^2\label{qxx}
				\\ 
				&
				\Gamma_m=\hbar v_F\frac{q^m_x+i(k_y+m\omega/v_F)}{E+m\hbar\omega-\Delta_v}=s e^{i\varphi_m}
			\end{align} 
		\end{widetext}
		with the angle $\varphi_m= \tan^{-1} \left(\frac{k_y+m\omega/v_F}{q^m_x}\right)$.
		Note that from \eqref{qxx}, we can determine the corresponding energy to \eqref{psi2}. 
		In regions 1 and 3, we have pristine tungsten diselenide, and  therefore, the eigenspinors are given by \cite{Elaitouni2023,laser1,laser2}
		\begin{widetext}
			\begin{align}
				&\Psi^1(x,y,t)=\sum_{m,l-\infty}^{\infty}\left[\begin{pmatrix}
					1\\ \beta_m
				\end{pmatrix}\delta_{l,0}e^{i\tau k^m_xx}+r_m\begin{pmatrix}
					1\\ -\beta^*_m
				\end{pmatrix}e^{-i\tau k^m_xx}\right]e^{ik_yy}e^{-i(E+m\hbar\omega)t/\hbar}\delta_{m,l} \label{spI}\\
				&
				\Psi^3(x,y,t)=\sum_{m,l-\infty}^{\infty}t_m\begin{pmatrix}
					1\\ \beta_m
				\end{pmatrix}e^{i\tau k^m_xx}e^{ik_yy}e^{-i(E+m\hbar\omega)t/\hbar}\delta_{m,l}\label{spT}
			\end{align} 
		\end{widetext}
		such that  
		\begin{align}
			& k^2 =(E+m\hbar\omega-\Delta_c)(E+m\hbar\omega-\Delta_v)/(\hbar v_F)^2\label{E8}\\
			&(k^m_x)^2 = k^2-k_y^2\\
			&\beta_m=\hbar v_F\frac{k^m_x+ik_y}{E+m\hbar\omega-\Delta_v}= s' e^{i\phi_m}  
		\end{align}
		with the incident angle %$\theta_m= \tan^{-1} \frac{k_y}{k^m_x}$.
		$\phi_m= \tan^{-1} \frac{k_y}{k^m_x}$ and  $ m=0$  we set $\phi_0 = \phi$. %with the angle of the $m$-th sideband given by $\theta_m = \tan^{-1}\!\left(\dfrac{k_y}{k^m_x}\right)$. It is important to note that the physical incident angle of the incoming fermion beam corresponds to the central band ($m = 0$). Throughout our numerical results and figures, we denote this initial incident angle simply as $\phi$ (where $\phi \equiv \theta_0$).} 
	
	The time-periodic laser field couples an incident carrier of energy $E$ to laser-dressed sidebands $E + l\hbar\omega$. Therefore, the transmission channels discussed here correspond to the Floquet sidebands, and their contributions are quantified by the sideband-resolved probabilities $T_l = |t_l|^2$. The laser parameter $\alpha = A_0/\omega$ enters through the Bessel coefficients $J_{l-m}(\alpha)$, which redistribute the transmission 
		weight among different sidebands. For each sideband, transport inside region 2 is governed by the longitudinal wave vector $q_x^m$ in \eqref{qxx}: a real $q_x^m$ corresponds to a propagating mode, whereas an imaginary $q_x^m$ corresponds to an evanescent mode that decays across the barrier. Hence, the reduction of total transmission at larger $\alpha$ or larger $D$ 
		results from both sideband redistribution and interference between forward and backward Floquet components inside the irradiated barrier.
	In the following, we apply the theoretical framework developed above to investigate the transmission and conductance properties of monolayer WSe$_2$. Using the derived Hamiltonian and the corresponding wave functions in each region, we compute the spin- and valley-resolved transmission probabilities by enforcing appropriate boundary conditions at the interfaces. These results are then employed within the Landauer–Büttiker formalism to evaluate the conductance. Through this analysis, we explore how strain, electrostatic potential, and intrinsic material parameters influence the transport behavior and the resulting spin- and valley-dependent responses in WSe$_2$.
	
	\section{Numerical results}\label{num}

	We numerically explore the transmission probability and conductance that were derived theoretically in Appendix \ref{Tran} and Appendix \ref{Cond}, focusing on the case of laser-assisted tunneling through a static tungsten diselenide (WSe$_2$) barrier. Based on the analytical framework established in those appendices, we examine how an external laser field modifies the tunneling process by implementing the corresponding expressions. Specifically, we aim to understand the impact of the interaction between the electromagnetic field and the electronic structure of WSe$_2$ on the transmission characteristics and resulting conductance.
	Through numerical analysis, we can view and understand how the system behaves in response to different physical conditions, such as field strength and energy variations. The system provides a visual demonstration of how photon-assisted processes allow particles to traverse the barrier. This research bridges the gap between mathematical proofs and their practical applications by offering an accurate explanation of laser-based movement in WSe2 structures.

	\begin{figure*}[ht!]
		\centering
		\subfloat[$\alpha=1$]{\includegraphics[scale=0.26]{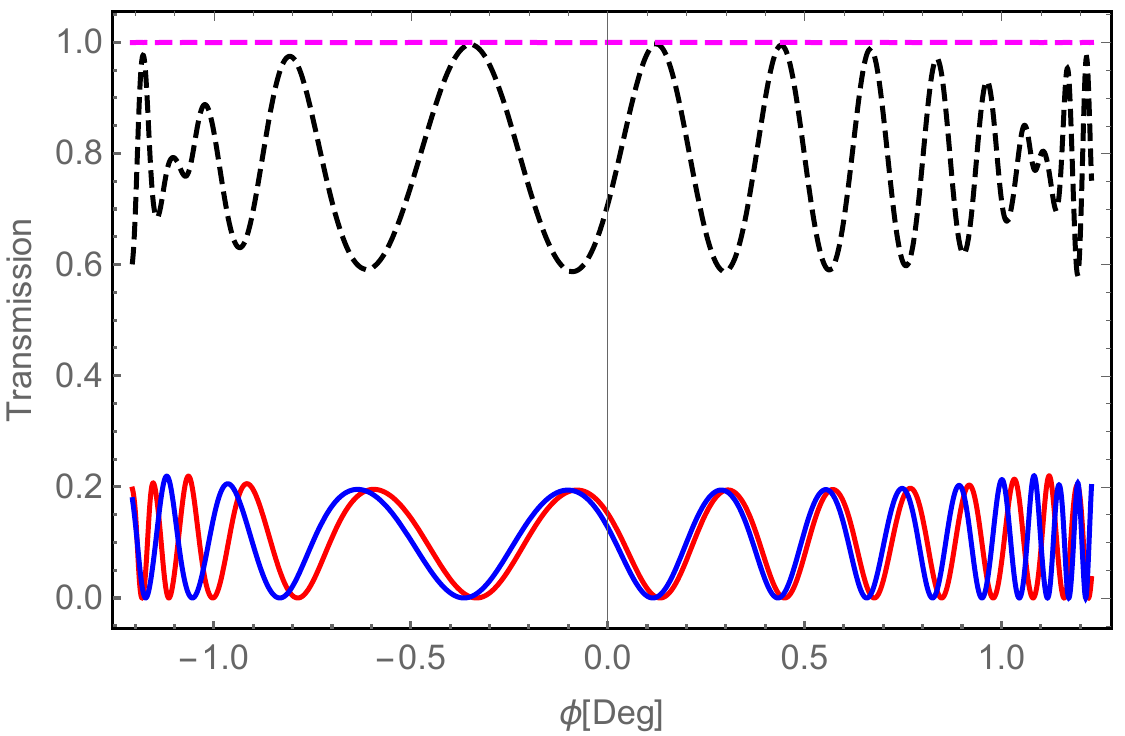}\label{phia}}
		\qquad
		\subfloat[ $\alpha=2$]{\includegraphics[scale=0.26]{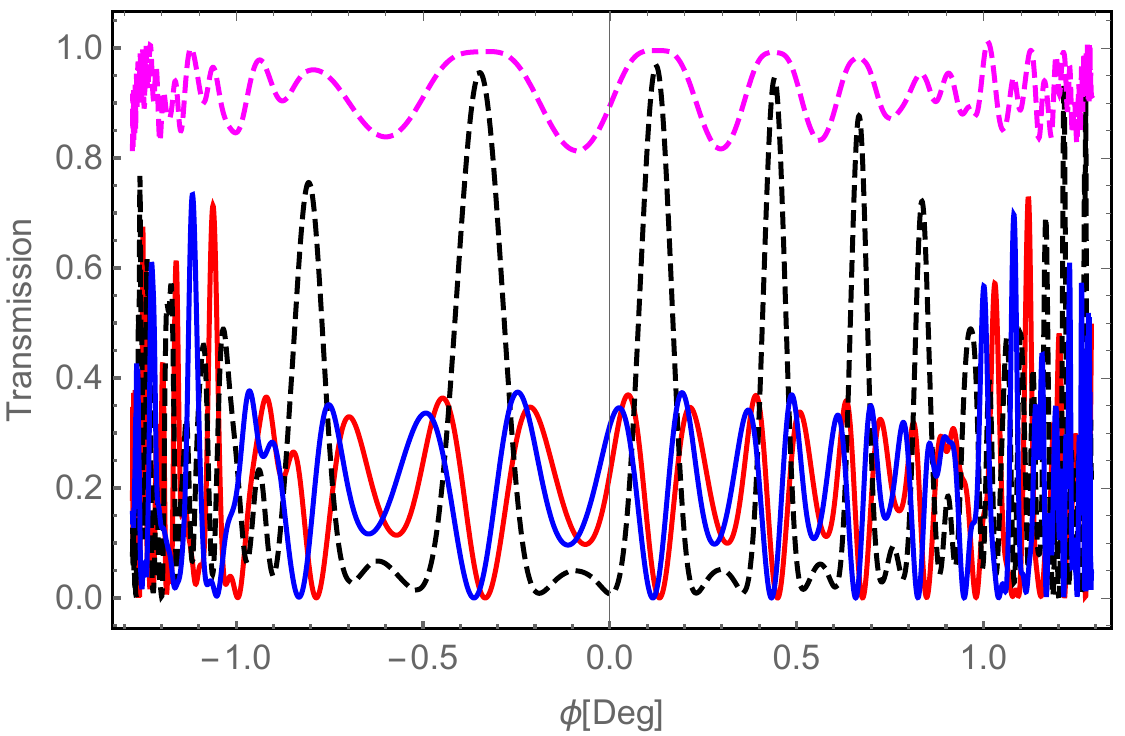}\label{phib}} \qquad
		\subfloat[$\alpha=3$]{\includegraphics[scale=0.26]{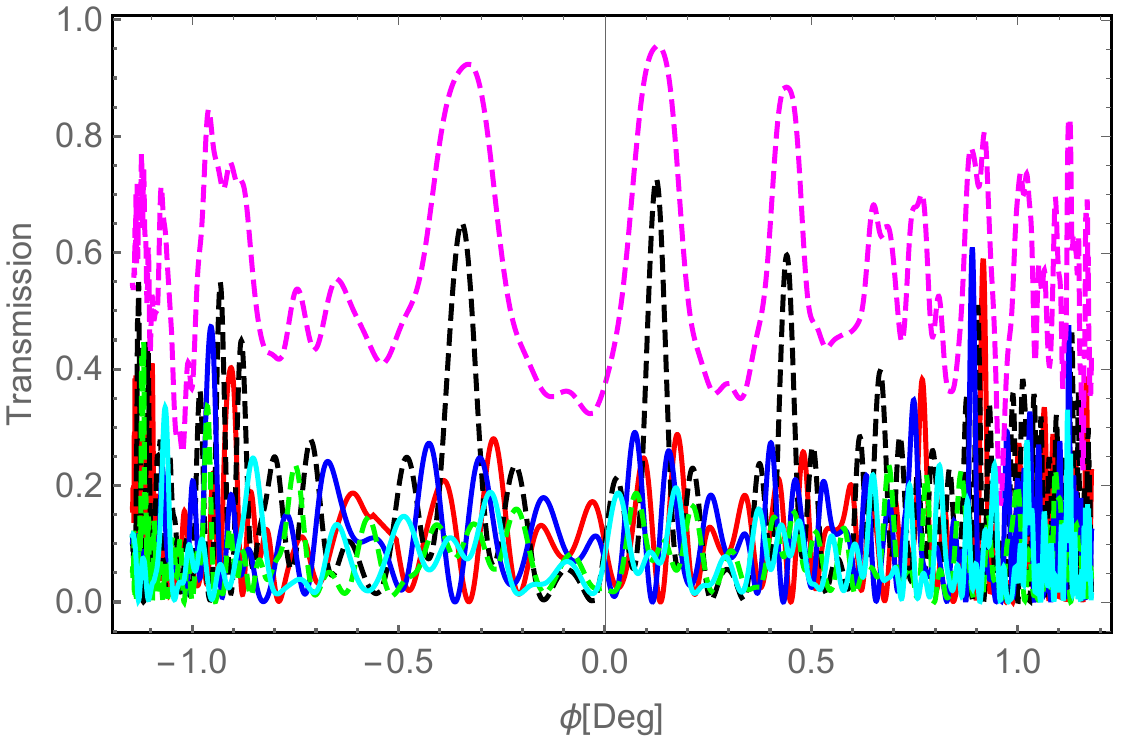}\label{phic}}
		\subfloat{\includegraphics[height=2.5cm]{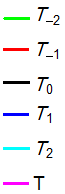}}
		\caption{Transmission as a function of incident angle $\phi$ for $D=150$ nm, $E=1$ eV, $V_0=3$ eV and $\omega=12.5\times 10^{13}$ Hz.}\label{phi}
	\end{figure*}

	\begin{figure*}[ht]
		\centering
		\subfloat[$D=150$ nm, $E=1.5$ eV]{\includegraphics[scale=0.23]{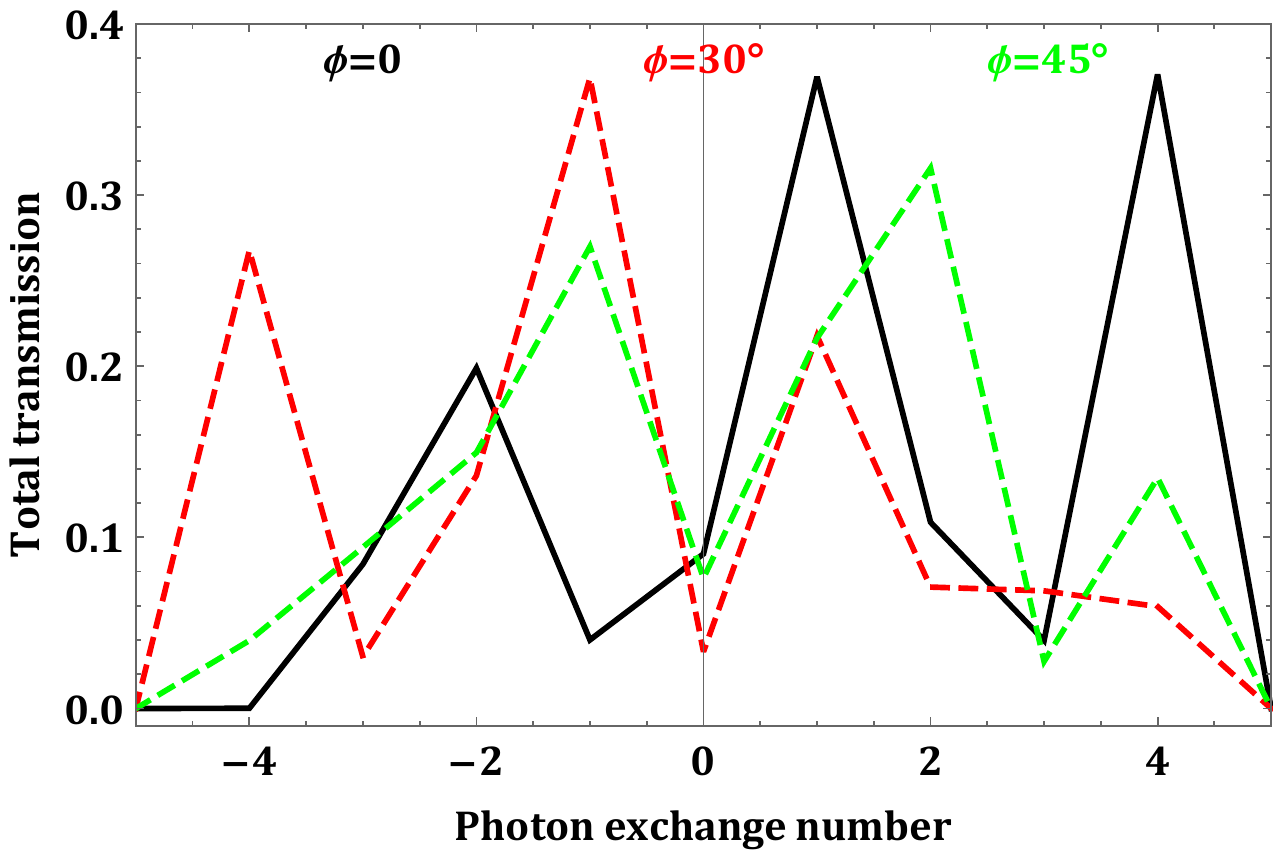}\label{photonphi}}\qquad
		\subfloat[ $\phi=0$, $E=1$ eV]{\includegraphics[scale=0.23]{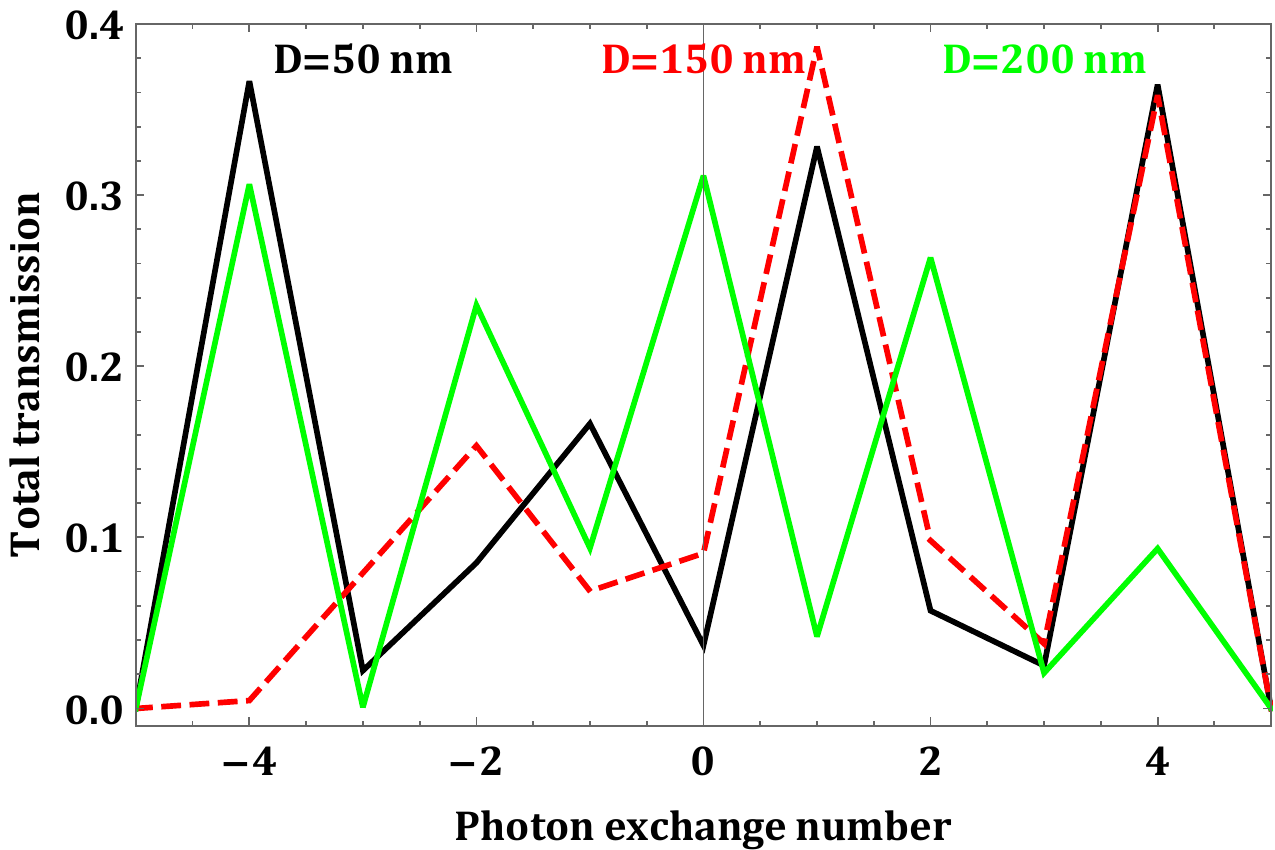}\label{photonD}}\qquad
		\subfloat[$\phi=30°$, $D=150$ nm]{\includegraphics[scale=0.23]{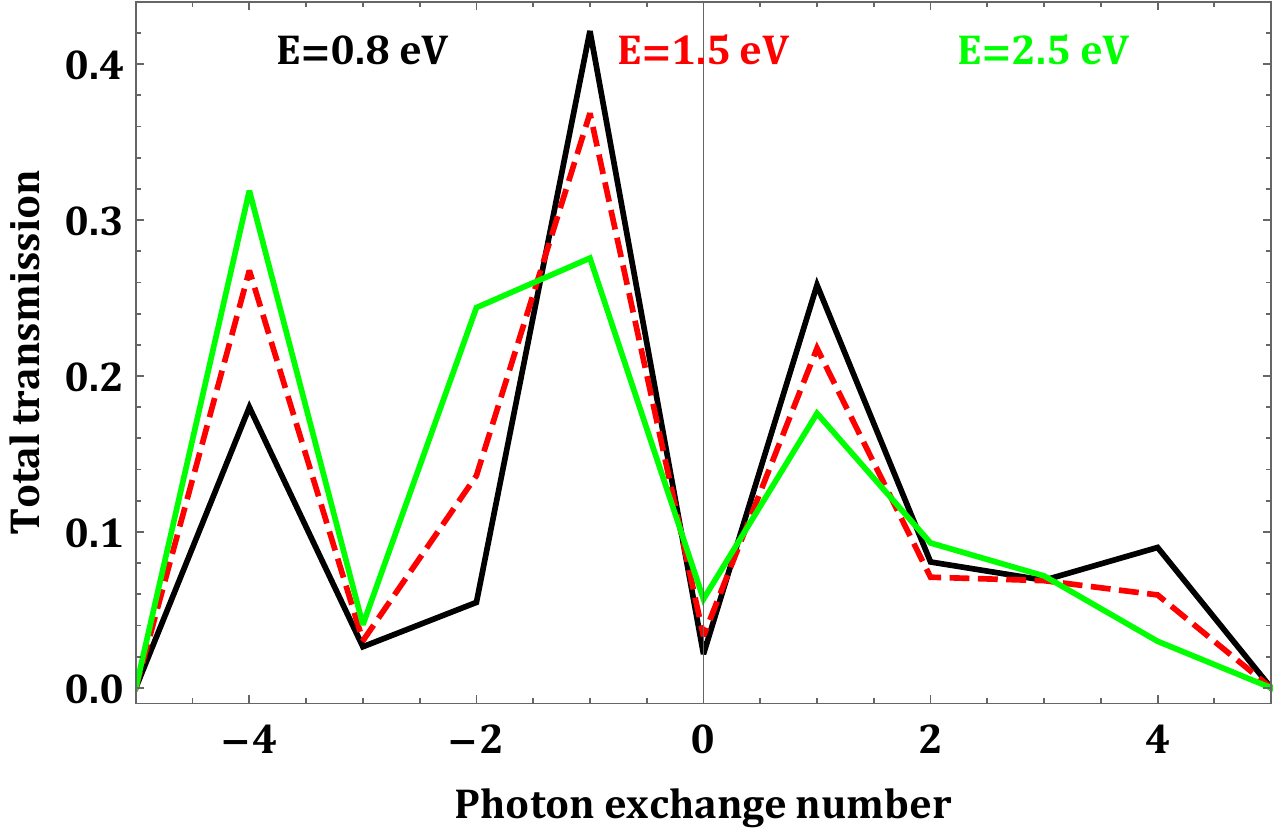}\label{photonE}}
		\caption{Transmission  as a function of the number of photon exchanged for $\omega=12.5 \times 10^{13}$ Hz, and $\alpha=4$.}\label{photon}
	\end{figure*}
	
	Figure~\ref{phi} illustrates the transmission  as a function of the  incident angle $\phi$ for an energy $E = 0.95$ eV,  barrier height $V_0 = 3$ eV and width $D = 150$ nm. The transmission exhibits an oscillatory behavior, dominated by the channel without photon exchange corresponding to the central band. In Fig.~\ref{phia}, obtained for a less intense laser ($\alpha = 1$), the condition $N < \alpha$ implies that only the central band and single-photon absorption or emission channels appear. The transmission without photon exchange is the most dominant, while absorption and emission contributions are nearly equal. The total transmission is perfect for all angles, indicating that fermions cross the barrier with a very low probability of photon exchange. 
	In Fig.~\ref{phib}, for a laser parameter $\alpha = 2$, the total transmission decreases and oscillates around unity. The transmission through the central band vanishes for several incidence angles, while absorption and emission channels, nearly equal, exceed $40\%$. Depending on the incidence angle, transmission occurs either exclusively with photon exchange or exclusively without photon exchange.  Finally, Fig.~\ref{phic}, corresponding to $\alpha = 3$, shows the emergence of new transmission channels involving two-photon exchange. Despite the opening of these additional channels, the total transmission decreases, highlighting the effect of laser irradiation on fermion dynamics. It turns out that laser irradiation is vital for managing fermion motion and choosing the transmission channel. When photons are exchanged between the barrier and fermions, their energy changes. This energy shift confines them to new levels, a phenomenon called the Stark effect \cite{Stark}. As the laser parameter increases, the suppression of the Klein tunneling effect is evident, and the transmission is no longer perfect for the majority of incident angles.
	
	\begin{figure*}[t]
		\centering
		\subfloat[$\phi=0$]{\includegraphics[scale=0.37]{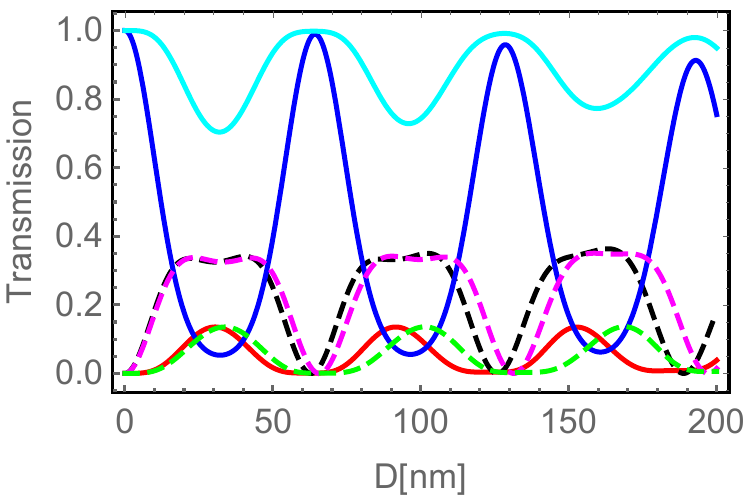}\label{TDa}} \qquad
		\subfloat[$\phi=\pi/10$]{\includegraphics[scale=0.37]{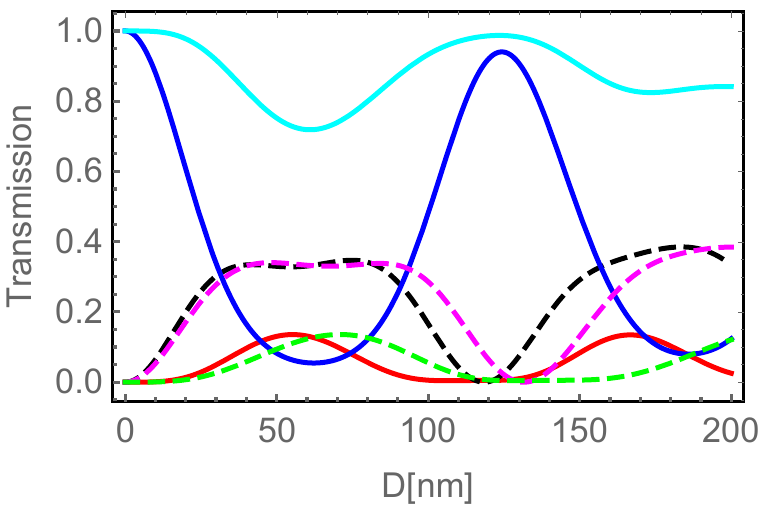}\label{TDb}}\qquad
		\subfloat[$\phi=\pi/3$]{\includegraphics[scale=0.37]{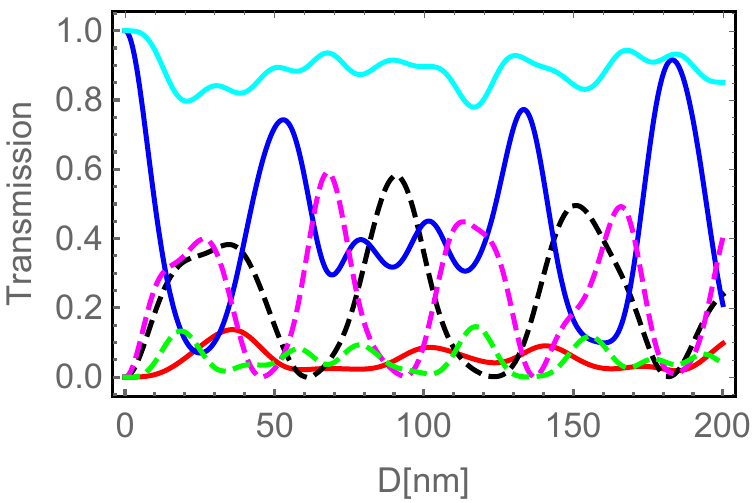}\label{TDc}}
		\subfloat{\includegraphics[height=2.4cm]{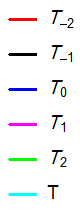}}
		\caption{Total transmission and transmissions of the first five side bands ($l=0,\pm 1,\pm 2$) as a function of barrier width for
			$\alpha=2$, $\omega=12.5\times 10^{13}$ Hz, $E=0.95$ eV, $V_0=1$ eV, and 
			three incident angles. }\label{TD}
	\end{figure*}
	
	\begin{figure*}[t]
		\centering
		\subfloat[$D=10$ nm]{\includegraphics[scale=0.37]{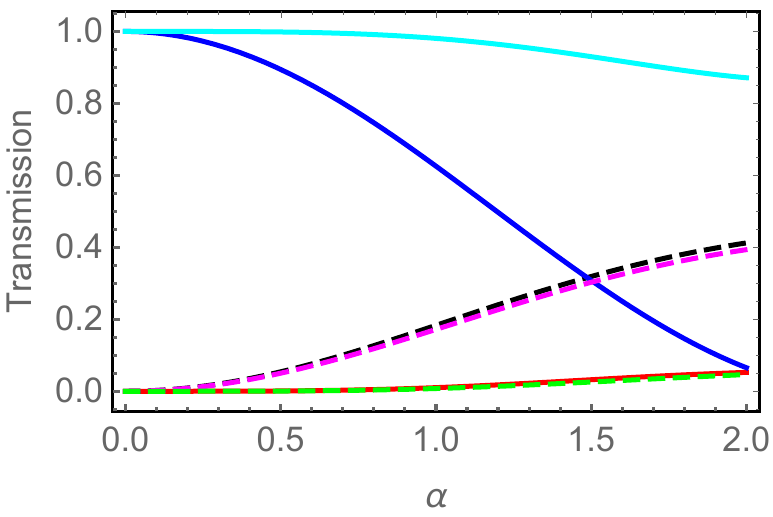}\label{alphaa}} \qquad
		\subfloat[$D=100$ nm]{\includegraphics[scale=0.37]{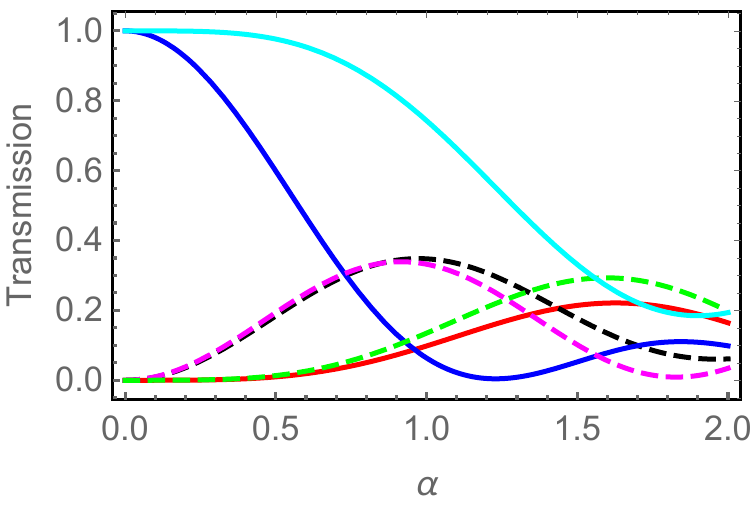}\label{alphab}}\qquad
		\subfloat[$D=200$nm]{\includegraphics[scale=0.37]{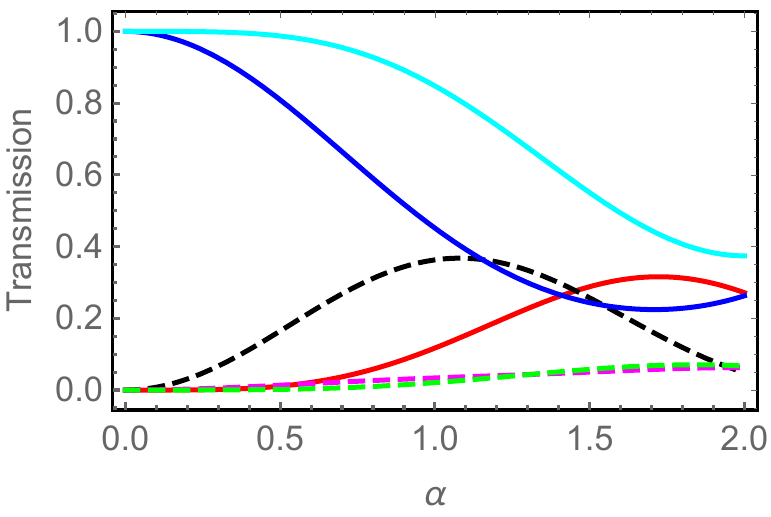}\label{alphac}}
		\subfloat{\includegraphics[height=2.4cm]{legend.png}}
		\caption{Total transmission and transmissions of the first 5 side bands ($l=0,\pm 1,\pm 2$) as a function of laser parameter $\alpha$ for 
			$\omega=12.5\times 10^{13}$ Hz, $E=0.95$ eV,  $V_0=1$ eV, and
			three different barrier widths.}\label{alpha}
	\end{figure*}

	For \( \alpha = 4 \) in Figure~\ref{photon}, we show how transmission varies with the number of photons exchanged with the barrier. This exchange occurs either by emission or absorption. Meanwhile, Fig.~\ref{photonphi} presents transmission  for three different incident angles. For normal incidence ($\phi = 0$) and for oblique incidence at $\phi = 30^\circ$, transmission is more likely via the emission of one or four photons. At an angle of $\phi = 45^\circ$, transmission is dominated by the exchange of two photons. Fig. \ref{photonD} represents the contribution of transmission channels at normal incidence for three barrier widths. When the width is small ($D = 50$ nm), four-photon exchange dominates through either absorption or emission. Increasing $D$ to 150 nm changes the behavior because either one-photon or four-photon emission emerges as the dominant process. For a very wide barrier ($D = 200$ nm), transmission without photon exchange becomes the most probable. Fig. \ref{photonE} presents the contribution of transmission channels for an oblique incidence of $\phi = 30^\circ$. In this case, transmission is more likely through photon absorption than emission. Therefore, we can conclude that the contribution of transmission channels varies with the incidence angle, the incident energy, and the laser parameter. This allows transmission to be guided by precisely adjusting the system parameters. This discovery opens the way to the design of a new generation of devices based on transmission channel selection.

	Figure~\ref{TD} displays three transmission types,  including total transmission, transmission without photon exchange, and transmission with one or two photon emission or absorption. The results show these three transmission types as functions of barrier width $D$ for  $\alpha =2$ and $E =0.95$ eV, and $V_0=1$ eV.
	At  normal incidence ($\phi=0$) in
	Fig. \ref{TDa}, it is observed that the total transmission is perfect for specific values of $D$, a phenomenon known as the Klein tunneling effect \cite{klien1,klien2,klienexp}. This effect represents a major issue in two-dimensional materials: even in the presence of a band gap, it hinders the use of most 2D materials such as graphene and molybdenum disulfide MoS$_2$ in advanced technologies. The transmission without photon exchange remains dominant, although other channels exist. Perfect transmission occurs when the lateral transmission vanishes. The exchange of a single photon during transmission is more probable than the exchange of two photons. 
	For $\phi=\pi/10$ in Fig. \ref{TDb}, the range of cancellation of lateral transmissions becomes larger and the points with perfect transmission decrease, therefore diminishing the proposed tunneling phenomena. Fig. \ref{TDc} indicates that perfect transmission is no longer achieved for an oblique incidence of $\phi = 30^\circ$, then the tunneling effect is fully suppressed. In particular, transmissions with photon exchange are enhanced compared with those without photon exchange for certain barrier widths.
	We see that widening the barrier enables fermions to interact with the laser irradiation, leading to the emergence of transmission channels involving photon exchange. This interaction alters the energy levels of the fermions, potentially trapping them between these new states, and increases the likelihood of destructive interference due to a reduced probability of presence. Altogether, these effects account for the suppression of the Klein tunneling effect.

	% Figure~\ref{alpha} illustrates the transmission as a function of the laser parameter for three barrier widths. For a narrow barrier of $D=10$ nm (Fig. \ref{alphaa}), the transmission without photon exchange decreases, while the transmission with the exchange of a single photon increases; the total transmission decreases only slightly. For a wider barrier of $D=100$ nm (Fig. \ref{alphab}), the transmission without photon exchange decreases rapidly as the transmission with photon exchange increases. In this case, the transmission through the exchange of two photons becomes as probable as that with the exchange of a single photon. The Klein tunneling effect is suppressed by increasing $\alpha$. For a very wide barrier of $D=200$ nm (Fig. \ref{alphac}), transmission through photon absorption becomes more probable, while transmission through photon emission is nearly zero. In this situation, the large barrier gives fermions a greater chance of destructive interference, even when their energy is high. In conclusion, increasing the laser field parameter allows certain transmission modes to be activated or deactivated. Photon exchange between the barrier and the fermions can reduce their ability to cross the barrier, thus providing an additional mechanism for controlling transmission channels.

	Figure~\ref{alpha} shows how transmission varies with laser parameters across three different barrier widths. The narrow barrier maintains a width of $D=10$ nm (Fig. \ref{alphaa}), which results in decreased transmission without photon exchange while transmission with one photon exchange increases, and total transmission experiences only a minor decline. The wider barrier maintains a width of $D=100$ nm (Fig. \ref{alphab}), which results in a rapid decrease of transmission without photon exchange while transmission with photon exchange shows an upward trend. 
	In this case, the transmission through the exchange of two photons becomes as probable as that with the exchange of a single photon. The Klein tunneling effect is suppressed by increasing $\alpha$. For a very wide barrier of $D=200$ nm (Fig. \ref{alphac}), transmission through photon absorption becomes more probable, while transmission through photon emission is nearly zero. In this situation, the large barrier gives fermions a greater chance of destructive interference, even when their energy is high. In conclusion, increasing the laser field parameter allows certain transmission modes to be activated or deactivated. Photon exchange between the barrier and the fermions can reduce their ability to cross the barrier, thus providing an additional mechanism for controlling transmission channels.

		In principle, the calculation of the conductance involves a summation over an infinite number of angular momentum modes as well as Floquet sidebands, which makes an exact numerical treatment impractical. To ensure the reliability of our results, we have carefully examined the convergence of the conductance with respect to both contributions. We find that the transmission probability decreases rapidly with increasing angular momentum index $|l|$, as higher-order modes are suppressed by the effective centrifugal barrier. In particular, modes with $|l|>2$ provide negligible contributions over the entire range of parameters considered. Consequently, restricting the summation to the first five modes, $l=0,\pm1,\pm2$, is sufficient to obtain fully converged results. 
		Furthermore, we have analyzed the convergence with respect to the number of Floquet sidebands associated with photon-assisted processes. Our calculations show that processes involving the exchange of more than two photons have vanishingly small amplitudes and do not affect the conductance. Therefore, truncating both the angular momentum modes and the Floquet sidebands is well justified and ensures an accurate and efficient numerical evaluation of the transport properties.

	% Figure~\ref{Galpha} illustrates the variation of conductance as a function of the laser irradiation parameter $\alpha$ for three distinct barrier widths. A monotonic decrease in conductance is observed as $\alpha$ increases, highlighting the influence of laser irradiation on fermion mobility.
	%{\bb In the two forthcoming figures, we present the conductance as a function of $\alpha$
		%and $D$. The use of all transmission modes is impossible since they are unlimited. As shown in Fig.~\ref{phi} and Fig.~\ref{TD}, the contribution of the lateral bands to the transmission is very weak, which justifies restricting the integration to the first five modes, namely $l=0,\pm 1, \pm 2$, in order to facilitate the numerical calculation on our machine. Moreover, the transmission involving the exchange of more than two photons is negligible, implying that the corresponding conductance is also less significant for these modes.}

\begin{figure}[ht]
		\centering
		\includegraphics[scale=0.55]{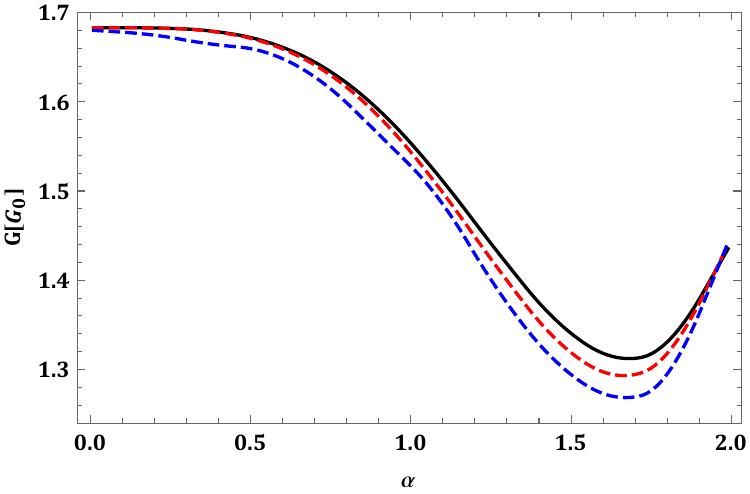}
		\caption{The conductance as a function of the laser parameter $\alpha$
			for $V_0=2$ eV, $E=1.2$ eV, $\omega=12.5\times 10^{13}$ Hz, and
			three different barrier widths $D=50$ nm (black line), $D=150$ nm (red dashed line), $D=250$ nm (blue dashed line).}\label{Galpha}
	\end{figure}

	Figure~\ref{Galpha} shows how conductance changes with different values of laser irradiation parameter $\alpha$ for three different barrier width measurements. The conductance shows a continuous decline, which correlates with rising values of $\alpha$ because laser irradiation affects fermion mobility.
	This effect arises from photon exchange between the barrier and the fermions, which modifies their energy levels and can trap them between newly quantized states—a phenomenon analogous to the Stark effect—leading to a subsequent reduction in the overall current across the barrier \cite{Stark}. Note that, conductance is inversely proportional to the barrier width. Then, wider barriers provide fermions with a longer interaction path with the field, thus increasing the probability of destructive interference between fermion states inside the barrier \cite{Nandy2019,Dajka2024}.  These findings demonstrate that the laser parameter serves as an effective external mechanism for tuning conductance in WSe$_2$-based structures. This approach offers a promising route toward optically controllable electronic devices, enabling dynamic modulation of transport properties without altering the structural configuration.

	% Figure~\ref{GD} exhibits the condusctance as a function of  the barrier width $D$ for $V_0 = 2$ eV,  $E = 1.2$ eV, and  $\omega = 12.5 \times 10^{13}$ Hz. For the three examined laser intensities ($\alpha = 1, 1.5,$ and 2), conductance exhibits a decreasing trend as $D$ increases from the nanoscale up to around $80$ nm, highlighting the significant impact of barrier geometry on charge transport. 
	% This attenuation can be explained by the prolonged dwell time of fermions within the irradiated region: a wider barrier enhances the interaction between fermions and the laser field, promoting photon absorption and emission processes that redistribute transmission probabilities across different energy channels. Consequently, the fraction of fermions crossing the structure without photon exchange decreases with increasing laser field intensity, leading to a net reduction in conductance. From a device engineering perspective, these results indicate that compact barrier designs with minimal $D$ values are optimal for achieving high conductance, whereas wider barriers combined with intense laser fields provide an effective strategy for significant current suppression in WSe$_2$-based nanostructures.
	
	Figure~\ref{GD} displays how conductance varies according to barrier width $D$ while keeping $V_0 = 2$ eV, $E = 1.2$ eV, and $\omega = 12.5 \times 10^{13}$ Hz as constant values. Three different $\alpha$ values (1,  1.5,  2) create three distinct curves. The study demonstrates that conductance decreases progressively when barrier width increases from nanometer range to 80 nm. The study demonstrates that charge transport exhibits high sensitivity toward the barrier geometric dimensions.
	This phenomenon
	can be explained through simple physical principles. When the barrier width expands, fermions spend extended periods inside the irradiated area. The longer they stay there, the more they will interact with the laser field. The stronger interaction leads to higher chances of photon absorption and emission events. The redistribution of transmission probability happens through multiple photon-assisted pathways, which results in decreased overall conductance.
	The photon-assisted processes become more intense when laser fields reach higher levels because this effect occurs at higher $\alpha$ values. The stronger laser light creates conditions, which lead to multiple photon exchanges while it disrupts the coherent transmission pathway through the barrier. The increase in laser field strength results in more substantial conductance reductions.
	The results offer device designers helpful information about their design work. Designers who want to achieve high conductance should select narrow barriers that use small $D$ values because this choice minimizes carrier–field interactions. The combination of wide barriers with strong laser irradiation offers an effective method to control current flow, which benefits applications including optical switching and adjustable transport management in WSe$_2$-based nanodevices.

	\begin{figure}[ht]
		\centering
		\includegraphics[scale=0.55]{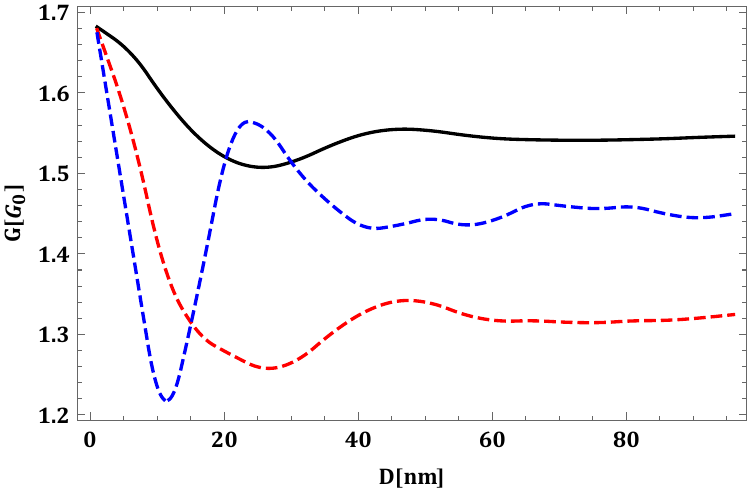}\label{GD}
		\caption{Conductance as a function of  barrier width $D$ for 
			$V_0=2$ eV, $E=1.2$ eV,  $\omega=12.5\times 10^{13}$ Hz, and 
			three different parameters  $\alpha=1$ (black line), $\alpha=1.5$ (red dashed line), $\alpha=2$ (blue dashed line).}\label{GD}
	\end{figure}
	
	These results demonstrate the rich tunability of the conductance in the 
		WSe$_2$ barrier. It is worth noting that, unlike graphene, which is 
		gapless and has negligible spin-orbit coupling,  the WSe$_2$ barrier 
		possesses intrinsic and fixed values of both the band gap and spin-orbit 
		coupling strength, determined by its crystal structure and atomic composition. 
		The tunability of the conductance is therefore fully controlled by the laser 
		parameter $\alpha$ and the barrier width $D$, as shown in Figs.~\ref{Galpha} 
		and~\ref{GD}, providing continuous control over the range 
		$\sim 1.2\,G_0$ to $\sim 1.7\,G_0$.
	
	\section{Conclusion}\label{concl}
	
	This work presents a theoretical study of the effect of laser irradiation on transmission and conductance across a static barrier in tungsten diselenide WSe$_2$. The periodicity of the Hamiltonian under laser irradiation requires the use of Floquet theory to determine the wave functions in the three regions of the structure. The continuity of the wave function across these regions allows the calculation of transmission and reflection coefficients. The presence of Floquet modes complicates the resolution of the system, which necessitates a matrix formalism. The resulting transfer matrix is of infinite order, requiring an approximation based on Bessel functions, where higher-order terms beyond $\alpha$ are neglected. The continuity equation, which relates current densities to state densities, is then used to determine the transmission and reflection probabilities. Finally, to connect microscopic processes with macroscopic observables, the Büttiker relation is employed to calculate the conductance $G$.

	Numerical analysis shows that laser irradiation plays a crucial role in controlling transmission: as the laser intensity increases, photon exchange between fermions and the barrier becomes more significant, leading to modifications of energy levels. This exchange redistributes the incident flux among Floquet sidebands; when some of these sidebands are evanescent inside the barrier, their amplitudes decay across region 2, leading a decrease in  the total transmission.
	Increasing the barrier width enhances the probability of photon exchange but may also reduce transmission due to destructive interference. Adjusting system parameters such as laser intensity, incidence energy, and incidence angle enables suppression of Klein tunneling and allows control over transmission channels—whether without photon exchange, with single-photon exchange, or with exchange of a specific number of photons. Conductance follows the transmission behavior with respect to system parameters: increasing laser intensity reduces conductance, as fewer fermions traverse the barrier due to both Stark effects and destructive interference.

		Our results open promising avenues for experimental realization of next-generation optoelectronic devices. In particular, they suggest the possibility of designing transistors in which the electronic current is governed by laser irradiation rather than conventional gate voltages. They also provide a platform for highly sensitive optical sensors capable of detecting variations in the electromagnetic field through controlled modulation of electronic transport. Moreover, the strong selectivity of the laser-induced transmission channels offers new opportunities for quantum filtering applications. Finally, these findings are highly relevant for spintronics and valleytronics, where they could enable device concepts based on the manipulation of spin and valley degrees of freedom.
	%	These results can be exploited experimentally for the development of innovative optoelectronic devices. They pave the way for the fabrication of transistors in which the current is controlled by laser irradiation rather than by a gate voltage. They also enable the design of optical sensors capable of detecting variations in the light field through the modulation of electronic transport. Furthermore, these findings open perspectives for quantum filtering, thanks to the selectivity of transmission channels induced by the applied laser. Finally, they are of major interest for spintronics and valleytronics, with potential applications in the realization of devices based on spin or valley degrees of freedom (polarization).}

%\vspace{5mm}
\appendix
\section{Transmission}\label{Tran}

%\subsubsection{Transmission}
% The continuity of the wave function across the three regions 
% makes it possible to determine the transmission and reflection coefficients. These coefficients are particularly important, as they allow the calculation of the total transmission and reflection, and are also used to evaluate the conductance. They are
% \begin{align}
	%    & \Psi_1(0,y,t)=\Psi_2(0,y,t)\\
	%    &\Psi_2(D,y,t)=\Psi_3(D,y,t)
	% \end{align}
% They yield four equations, each involving an infinite number of modes. In this case, the system can be written in matrix form to simplify the calculation

The continuity of the wave function across the three regions 
makes it possible to determine the transmission and reflection coefficients. These coefficients are particularly important, as they allow the calculation of the total transmission and reflection probabilities, and are also used to evaluate the conductance within the Landauer framework. They are imposed through the boundary conditions %$ \Psi^1(0,y,t)=\Psi^2(0,y,t), \Psi^2(D,y,t)=\Psi^3(D,y,t)$.
\begin{align}
	& \Psi^1(0,y,t)=\Psi^2(0,y,t)\\
	&\Psi^2(D,y,t)=\Psi^3(D,y,t).
\end{align}
The application of matching conditions at both interfaces produces four equations which describe the continuity of two spinor components at two locations, $x=0$ and $x=D$. As a result, we obtain
\begin{widetext}
	\begin{align}
		&\delta_{l,0}+r_l=\sum_{m=-\infty}^{\infty}\left[a_m+b_m\right]J_{l-m}(\alpha)\\
		&\delta_{l,0}\beta_l-r_l\beta^*_l=\sum_{m=-\infty}^{\infty}\left[a_m\Gamma_m-b_m\Gamma_m^*\right]J_{l-m}(\alpha)\\
		&t_le^{i\tau k_lD}+\mathbb{0}_le^{-i\tau k_lD}=\sum_{m=-\infty}^{\infty}\left[a_me^{i\tau q_mD}+b_me^{-i\tau q_mD}\right]J_{l-m}(\alpha)\\
		&t_l \beta_le^{i\tau k_lD}-\mathbb{0}_l\beta^*_le^{-i\tau k_lD}=\sum_{m=-\infty}^{\infty}\left[a_m\Gamma_m e^{i\tau q_mD}-b_m\Gamma^*_m e^{-i\tau q_mD}\right]J_{l-m}(\alpha)
	\end{align}
\end{widetext}
%
%
%The system contains infinite equations because time-periodic potential wave functions require Floquet mode expansion to describe their behavior through multiple photon sidebands. The unknown reflection and transmission amplitudes for each mode create an infinite system of linear equations, which need to be solved.
%The problem becomes solvable through truncating the infinite series by selecting a specific range of sidebands, which extends from $n=-N$ to $n=0$ and ends at $n=+N$. The resulting system of equations can then be rewritten in a compact matrix form, which considerably simplifies the analytical manipulation and numerical implementation. The solution of this matrix system provides reflection and transmission coefficients for all modes, which enables calculation of total transmission and subsequent determination of the structure's conductance. As a result, this system can be written in matrix form, where each coefficient represents a vector of infinite order
%\begin{widetext}
The system leads to an infinite set of equations due to the Floquet expansion of the time-periodic wave functions, which introduces multiple photon sidebands with unknown reflection and transmission amplitudes. The resulting system of equations can then be rewritten in a compact matrix form as
\begin{widetext}
	\begin{align}
		\mathbb{M}_1
		\begin{pmatrix}
			0\\
			\vdots\\
			1\\
			\vdots\\
			0\\
			r_{-l}\\
			r_{-l+1}\\
			\vdots
		\end{pmatrix}=\mathbb{M}_2
		\begin{pmatrix}
			a_{-m}\\
			a_{-m+1}\\
			\vdots\\
			b_{-m}\\
			b_{-m+1}\\
			\vdots
		\end{pmatrix}, \quad
		\mathbb{M}_3
		\begin{pmatrix}
			t_{-m}\\
			t_{-m+1}\\
			\vdots\\
			0\\
			0\\
			\vdots
		\end{pmatrix}=
		\mathbb{M}_4
		\begin{pmatrix}
			a_{-m}\\
			a_{-m+1}\\
			\vdots\\
			b_{-m}\\
			b_{-m+1}
			\vdots
		\end{pmatrix}
	\end{align}
\end{widetext}
where the involved matrices are given by
\begin{widetext}
	\begin{align}
		&	\mathbb{M}_1
		=\begin{pmatrix}
			\begin{matrix} 1 & 0& \cdots \\ 0&1&\cdots  \\ \vdots & \vdots & \vdots \end{matrix}&\begin{matrix} 1 & 0& \cdots \\ 0&1&\cdots  \\ \vdots & \vdots & \vdots \end{matrix}\\
			\begin{matrix} \beta_{-l} & 0& \cdots \\ 0&\beta_{-l+1}&\cdots  \\ \vdots & \vdots & \vdots \end{matrix}&\begin{matrix} -\beta^*_{-l} & 0& \cdots \\ 0&-\beta^*_{-l+1}&\cdots  \\ \vdots & \vdots & \vdots \end{matrix}
		\end{pmatrix}, \quad
		\mathbb{M}_1
		=\begin{pmatrix}
			\begin{matrix} 1 & 0& \cdots \\ 0&1&\cdots  \\ \vdots & \vdots & \vdots \end{matrix}&\begin{matrix} 1 & 0& \cdots \\ 0&1&\cdots  \\ \vdots & \vdots & \vdots \end{matrix}\\
			\begin{matrix} \beta_{-l} & 0& \cdots \\ 0&\beta_{-l+1}&\cdots  \\ \vdots & \vdots & \vdots \end{matrix}&\begin{matrix} -\beta^*_{-l} & 0& \cdots \\ 0&-\beta^*_{-l+1}&\cdots  \\ \vdots & \vdots & \vdots \end{matrix}
		\end{pmatrix}	\\		
		&\mathbb{M}_2=\begin{pmatrix}
			\begin{matrix} J_{-l-m} & J_{-l+(1-m)}& \cdots \\ J_{-l+1-m}&J_{-l+1+(1-m)}&\cdots  \\ \vdots & \vdots & \vdots \end{matrix}&	\begin{matrix} J_{-l-m} & J_{-l+(1-m)}& \cdots \\ J_{-l+1-m}&J_{-l+1+(1-m)}&\cdots  \\ \vdots & \vdots & \vdots \end{matrix}\\
			\begin{matrix} J_{-l-m}\Gamma_m & J_{-l+(1-m)}\Gamma_{m-1}& \cdots \\ J_{-l+1-m}\Gamma_{m}&J_{-l+1+(1-m)}\Gamma_{m-1}&\cdots  \\ \vdots & \vdots & \vdots \end{matrix}&\begin{matrix} -J_{-l-m}\Gamma^*_m & -J_{-l+(1-m)}\Gamma^*_{m-1}& \cdots \\ -J_{-l+1-m}\Gamma^*_{m}&-J_{-l+1+(1-m)}\Gamma^*_{m-1}&\cdots  \\ \vdots & \vdots & \vdots \end{matrix}
		\end{pmatrix}\\
		&\mathbb{M}_3=	\begin{pmatrix}
			\begin{matrix} 	e^{i\tau k_{-l}D} & 0& \cdots \\ 0&e^{i\tau k_{1-l}D}&\cdots  \\ \vdots & \vdots & \vdots \end{matrix}&\begin{matrix} e^{-i\tau k_{-l}D} & 0& \cdots \\ 0&e^{-i\tau k_{1-l}D}&\cdots  \\ \vdots & \vdots & \vdots \end{matrix}\\
			\begin{matrix} 	\beta_{-l}e^{i\tau k_{-l}D} & 0& \cdots \\ 0&\beta_{1-l}e^{i\tau k_{1-l}D}&\cdots  \\ \vdots & \vdots & \vdots \end{matrix}&\begin{matrix} -\beta^*_{l}e^{-i\tau k_{-l}D} & 0& \cdots \\ 0&-\beta^*_{1-l}e^{-i\tau k_{1-l}D}&\cdots  \\ \vdots & \vdots & \vdots \end{matrix}
		\end{pmatrix}\\
		&\mathbb{M}_4=\begin{pmatrix}
			\begin{matrix} J_{-l-m}e^{i\tau k_{-l}D} & J_{-l+(1-m)}e^{i\tau k_{-l}D}& \cdots \\ J_{-l+1-m}e^{i\tau k_{1-l}D}&J_{-l+1+(1-m)}e^{i\tau k_{1-l}D}&\cdots  \\ \vdots & \vdots & \vdots \end{matrix}&	\begin{matrix} J_{-l-m}e^{i\tau k_{-l}D} & J_{-l+(1-m)}e^{i\tau k_{-l}D}& \cdots \\ J_{-l+1-m}e^{i\tau k_{1-l}D}&J_{-l+1+(1-m)}e^{i\tau k_{1-l}D}&\cdots  \\ \vdots & \vdots & \vdots \end{matrix}\\
			\begin{matrix} J_{-l-m}e^{i\tau k_{-l}D}\Gamma_m & J_{-l+(1-m)}e^{i\tau k_{-l}D}\Gamma_{m-1}& \cdots \\ J_{-l+1-m}e^{i\tau k_{1-l}D}\Gamma_{m}&J_{-l+1+(1-m)}e^{i\tau k_{1-l}D}\Gamma_{m-1}&\cdots  \\ \vdots & \vdots & \vdots \end{matrix}&\begin{matrix} -J_{-l-m}e^{i\tau k_{-l}D}\Gamma^*_m & -J_{-l+(1-m)}e^{i\tau k_{-l}D}\Gamma^*_{m-1}& \cdots \\ -J_{-l+1-m}e^{i\tau k_{1-l}D}\Gamma^*_{m}&-J_{-l+1+(1-m)}e^{i\tau k_{1-l}D}\Gamma^*_{m-1}&\cdots  \\ \vdots & \vdots & \vdots \end{matrix}
		\end{pmatrix}
	\end{align}
\end{widetext}
From this matrix formulation, one can easily express the transmission and reflection coefficients as

\begin{widetext}
	\begin{align}
		\begin{pmatrix}
			t_{-m}\\
			t_{-m+1}\\
			\vdots\\
			0\\
			0\\
			\vdots
		\end{pmatrix}=
		\mathbb{M}
		\begin{pmatrix}
			0\\
			\vdots\\
			1\\
			\vdots\\
			0\\
			r_{-l}\\
			r_{-l+1}\\
			\vdots
		\end{pmatrix}, \quad 
		%\end{align}
		%with 
		%\begin{align}
		\mathbb{M}=\mathbb{M}^\dagger_3\mathbb{M}_4\mathbb{M}^\dagger_2\mathbb{M}_1
		=\begin{pmatrix}
			\begin{matrix} M_{1,1} & M_{1,2}& \cdots &M_{1,2m+1}\\
				M_{2,1}&M_{2,2}&\cdots &\cdots 
				\\ \vdots & \vdots&\vdots & \vdots \\
				M_{2l+1,1}& \cdots &\cdots&M_{2l+1,2m+1}
			\end{matrix}
		\end{pmatrix}
	\end{align}
\end{widetext}
The problem of infinite order, impossible to solve analytically. So we can reduce the order to a finite number for values from $-N$ to $N$, with $N$ an integer depending on $A_0$. Then we neglect the Bessel terms $J_n(\alpha)$ of order $n$ greater than $\alpha$ \cite{laser1}. Finally, the transmission coefficient can be expressed as
the matrix element of $ \mathbb{M}$
\begin{align}
	t_l=M[N+l+1,N+1]
\end{align}
with $l$ ranges over the interval $[-N,N]$.

Transmission is determined by applying the continuity equation, which allows one to evaluate the incident $J^{i}$, reflected $J^{r}$, and transmitted $J^{t}$ current densities. This is
\begin{align}
	\frac{\partial \rho(x,y,t)}{\partial t} + \nabla \cdot \mathbf{j}(x,y,t) = 0
\end{align}
where  $\rho(x,y,t)=|\Psi(x,y,t)|^2$, and the components of $\mathbf{j}(x,y,t) $ are given by
%\begin{widetext}
\begin{align}
	&J^{i}_{0} = %v_F \, \Psi^{\dagger}_{i}(x,y,t) \, \tau \, \sigma_x \, \Psi_{i}(x,y,t)=
	v_F  \tau (\beta_0+\beta^*_0)\\
	&J^{r}_{l} % = v_F \, \Psi^{\dagger}_{r}(x,y,t) \, \tau \, \sigma_x \, \Psi_{r}(x,y,t)
	=v_F  \tau |r_l|^2 (\beta_l+\beta^*_l)\\
	&J^{t}_{l} % = v_F \, \Psi^{\dagger}_{t}(x,y,t) \, \tau \, \sigma_x \, \Psi_{t}(x,y,t)
	=v_F  \tau |t_l|^2 (\beta_l+\beta^*_l).
\end{align}
%\end{widetext}
From these current densities, the transmission and reflection probabilities are calculated according to the following expression:
\begin{align}
	T_l=\frac{|J^{t}_l|}{|J^{i}_0|}=|t_l|^2, \quad R_l=\frac{|J^{r}_l|}{|J^{i}_0|}=|r_l|^2.
\end{align}
Because of the Floquet subbands, the transmission also has an infinite number of modes. Transmission through the central band $T_0$ occurs without photon exchange between the barrier and the fermions, whereas transmission through the sidebands involves photon exchange $T_l$. In this case, the total transmission is the sum over all transmission channels
\begin{align}
	T=\sum_{l=-N}^{N}T_l.
\end{align}
The study of all modes is impossible; therefore, we restrict our analysis to the first transmission modes in addition to the transmission through the central band.

\vspace{5mm}

\section{Conductance}\label{Cond}
To connect microscopic quantities to macroscopic ones, the conductance is calculated. By definition, at zero temperature the conductance is given by the integral of the total transmission over $k_y$ \cite{conduct2}, at the same time, it corresponds to the average fermion flux across half of the Fermi surface \cite{butker, conduct1}.
\begin{align}
	G_\tau = \frac{G_0}{2\pi} \int_{-k_y^{\text{max}}}^{k_y^{\text{max}}} (T_{\uparrow\tau}(E, k_y)+T_{\downarrow\tau}(E, k_y))\, dk_y	
\end{align}
where \(G_0=\frac{2\pi e^2}{\hbar}\) is the unit of conductance, and \(k_y^{\text{max}}\) represents the maximum wave vector component along the \(y\)-direction. The connection between the transverse wave vector \(k_y\) and the incident angle \(\phi\) is used to express \(G\) as 
\begin{equation}
	G_\tau = \frac{G_0}{2\pi} \int_{-\phi^{\text{max}}}^{\phi^{\text{max}}} \left[T_{\uparrow\tau}(E, \phi)+T_{\downarrow\tau}(E, \phi)\right]\cos\phi \, d\phi
\end{equation}
and $\phi^{\text{max}}$ can be obtained from the relation 
$(k_y^{\text{max}} = K \sin(\phi^{\text{max}})$, with $K$ given by \eqref{E8}. This relation provides a connection between the energy $E$ and the maximum angle $\phi_0^{\text{max}}$ for the allowed transmission channels.

\end{document}